\documentclass[preprint,12pt]{elsarticle}
\usepackage{amsmath}
\usepackage[usenames]{color}
\definecolor{blue}{rgb}{0,0.08,0.45}

\usepackage{amssymb}
\newcommand{\mean}[1]{\langle#1\rangle}

\newcommand{\ud}[1]{#1^\dagger}

\journal{Superlattices and Microstructures}

\begin{document}

\begin{frontmatter}

%% Title, authors and addresses

%% use the tnoteref command within \title for footnotes;
%% use the tnotetext command for theassociated footnote;
%% use the fnref command within \author or \address for footnotes;
%% use the fntext command for theassociated footnote;
%% use the corref command within \author for corresponding author footnotes;
%% use the cortext command for theassociated footnote;
%% use the ead command for the email address,
%% and the form \ead[url] for the home page:
%% \title{Title\tnoteref{label1}}
%% \tnotetext[label1]{}
%% \author{Name\corref{cor1}\fnref{label2}}
%% \ead{email address}
%% \ead[url]{home page}
%% \fntext[label2]{}
%% \cortext[cor1]{}
%% \address{Address\fnref{label3}}
%% \fntext[label3]{}

  \title{Anticrossing in the PL spectrum of light-matter coupling
    under incoherent continuous pumping}

%% use optional labels to link authors explicitly to addresses:
%% \author[label1,label2]{}
%% \address[label1]{}
%% \address[label2]{}

\author[1]{A. Gonzalez-Tudela}
\author[2]{E. del Valle}
\author[1]{C. Tejedor}
\author[2]{F.P. Laussy}

\address[1]{F\'{\i}sica Te\'orica de la Materia Condensada,
  Universidad Aut\'onoma de Madrid, Spain.}

\address[2]{Physics and Astronomy, University of Southampton, U.K.}

\begin{abstract}
  We compare the observed splitting in the PL spectrum of a strongly
  coupled light-matter system, with the splitting of its dressed
  modes.  In the presence of non-negligible decoherence, the two may
  differ considerably. Whereas the dressed mode splitting has a simple
  expression, the observed splitting has no general analytical
  expression in terms of radicals of the system parameters.
\end{abstract}

\begin{keyword}
%% keywords here, in the form: keyword \sep keyword
  quantum dots \sep microcavities \sep strong coupling \sep PL
  spectrum
%% PACS codes here, in the form: 
\PACS 42.50.Ct\sep 78.67.Hc \sep 42.55.Sa \sep 32.70.Jz 
%% MSC codes here, in the form: \MSC code \sep code
%% or \MSC[2008] code \sep code (2000 is the default)
\end{keyword}

\end{frontmatter}

There is now a large number of reports of strong light-matter coupling
in zero dimensional semiconductor systems (one quantum dot in a
microcavity). The successful theoretical description of some of these
experiments relies on properly taking into account dissipation and
decoherence~\cite{qdth}. In these approaches, light-matter coupling is
described with the linear model Hamiltonian
$H=\omega_a\ud{a}a+(\omega_a-\Delta)\ud{b}b+g(\ud{a}b+a\ud{b})$
where~$a$ and~$b$ are the cavity photon and material excitation
(bosonic) field operators, respectively, with bare mode
energies~$\omega_a$ and~$\omega_a-\Delta$, coupled with strength~$g$.
The photoluminescence (PL) spectrum of the cavity emission, analyzed
in the steady state (SS), is greatly affected by decoherence. On the
one hand, the system loses photons and matter excitations at
rates~$\gamma_a$ and~$\gamma_b$, respectively. On the other hand, it
is driven by a cw off-resonant pumping or, equivalently, by a
continuous electronic injection in the wetting layer. This is
effectively modelled by two excitation rates, $P_a$ and~$P_b$. The
dynamics of the system is given by a standard master equation. In
Ref.~\cite{bosons}, we investigated in detail the SS dynamics and PL
spectrum of this system and obtained general analytical expressions
for the lineshapes:
\begin{equation}
 \label{eq:TueApr28150718GMT2009}
 S(\omega)=\frac{1}{2\pi}\Re\Big\{\frac{i-W}{\omega-\Omega_-}+ \frac{i+W}{\omega-\Omega_+}\Big\}=\Re\{A(\omega)\}\,,
\end{equation}
where~$A(\omega)$ is a complex function of the real
frequency~$\omega$, with:\\
\\
$\Omega_\pm=\omega_a-\Delta/2-i\Gamma_+\pm R \,\in \mathbb{C} \quad \text{(complex frequencies)}$\,,\\
$W=[\Gamma_-+i(\Delta/2+gD)]/R\,\in \mathbb{C} \quad \text{(dimensionless)}$\,,\\
$\Gamma_{\pm}=(\Gamma_a\pm\Gamma_b)/4\,\in \mathbb{R}\,,\quad \Gamma_{a,b}=\gamma_{a,b}-P_{a,b}\,\in \mathbb{R}\quad \text{(decoherence rates)}$\,,\\
$R=\sqrt{g^2-\left(\Gamma_-+i\Delta/2\right)^2}\,\in \mathbb{C}\quad\text{(half-Rabi frequency)}$\,,\\
$D=\frac{\mean{\ud{a}b}^\mathrm{SS}}{\mean{\ud{a}a}^\mathrm{SS}}=
\frac{\frac{g}2(\gamma_aP_b-\gamma_bP_a)(i\Gamma_+-\Delta/2)}
{g^2\Gamma_+(P_a+P_b)+P_a\Gamma_b(\Gamma_+^2+(\Delta/2)^2)}\,\in
\mathbb{C}\quad\text{(dimensionless)}$\,.\\
\\
With these definitions, the strong coupling (SC) regime is achieved
when the condition~$|\Gamma_-|<g$ is satisfied. In such a regime, the
so-called dressed modes (DM) appear at the
eigenfrequencies~$\Re\{\Omega_\pm\}$, overtaking bare modes (of the
weak coupling (WC) regime) at the frequencies~$\omega_a$ and
$\omega_b-\Delta$.

The observed spectrum, on the other hand, results from contributions
from the leading modes (bare in WC or dressed in SC), so its splitting
naturally does not correspond to that of the DM. Emission of each mode
is not Lorentzian but also has a dispersive part, that can result in
an increase or decrease of the apparent splitting, or even in its
appearance in weak coupling or disappearance in strong coupling. It is
therefore crucial, experimentally, to have an expression for the
observed splitting that can be directly compared with the observed
data.

It is possible to fit the lineshapes with the
expression~(\ref{eq:TueApr28150718GMT2009})~\cite{qdth}. In this text,
we focus on the alternative practice of fitting the maxima of these
spectra, that yield the characteristic crossing or anticrossing
behaviors. The advantages are in the simplicity (two lines rather than
a whole series of curves) and the robustness against noise (a
parasitic second dot shifts only slightly the maxima of the
strongly-coupled system whereas its influence on the whole lineshape
can be much more deleterious for the fitting, calling for outliers or
similar techniques). We need, however, to consider the observed
splitting, not the dressed mode splitting. We shall see however that
this approach is hindered by the Abel-Ruffini theorem that prohibits
solutions in terms of radicals of the parameters.

For this purpose, we consider the equation~$dS/d\omega=0$, that gives
the local extrema of the spectrum. There exists either one or three
real solutions, corresponding to the singlet or doublet, respectively.
Using~$dS/d\omega=\Re\{dA/d\omega\}$, that holds thanks to~$\omega \in
\mathbb{R}$, we obtain the factorization:
\begin{equation}
  \label{eq:TueApr28184248GMT2009}
  \frac{dS(\omega)}{d\omega}=\frac{1}{\pi}\frac{\Im\{(\omega-\omega_-)(\omega-\omega_+)(\omega-\Omega_-^*)^2(\omega-\Omega_+^*)^2\}}{|\omega-\Omega_-|^4|\omega-\Omega_+|^4}=0\,,
\end{equation}
with~$\omega_\pm=\Big[\Omega_+-\Omega_-+iW(\Omega_++\Omega_-)\pm
i\sqrt{(1+W^2)(\Omega_+-\Omega_-)^2}\Big]/2\in\mathbb{C}$. Despite the
simple form of Eq.~(\ref{eq:TueApr28184248GMT2009}), none of the
roots~$\{\omega_{\pm},\Omega_\pm^*\}$ are solutions of it, since they
are not real.  All the solutions must be found taking the imaginary
part in Eq.~(\ref{eq:TueApr28184248GMT2009}). This yields a quintic
equation. It is well known that a quintic cannot be solved in general
in terms of radicals of its coefficients.  This is the case of
Eq.~(\ref{eq:TueApr28184248GMT2009}) for a general set of
parameters. The solutions can still be presented in terms of special
functions, such as the Jacobi theta functions, but such elaborate
methods do not serve our objective of providing formulas that can be
directly used to analyze the experimental data. Therefore, in the most
general case, we prefer to solve numerically
Eq.~(\ref{eq:TueApr28184248GMT2009}), exploiting the simplicity of the
equation rather than the complexity of its solutions.

\begin{figure}[tbp]
  \centering
  \includegraphics[width=\linewidth,angle=0]{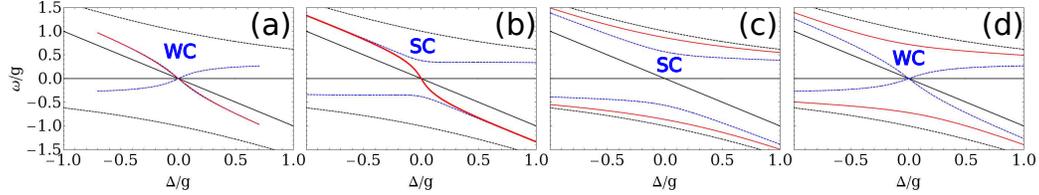}
  \caption{Dressed modes (blue) and observed PL splitting (red) as a
    function of detuning. We fix~$\gamma_b=0.1g$ and~$P_a=0.5g$.
    (a)~$\gamma_a=3.8g$ and~$P_b=g$, (b) $\gamma_a=3.8g$
    and~$P_b=0.5g$, (c)~$\gamma_a=3.8g$ and~$P_b=0.1g$, and
    (d)~$\gamma_a=4.8g$ and~$P_b=0.1g$.}
  \label{fig:MonFeb16000429CET2009}
\end{figure}

In Fig.~\ref{fig:MonFeb16000429CET2009}, we plot, as a function of
detuning: in solid black, the bare modes; in dashed black, the DM in
the absence of pump and decay; in dashed blue, the system DM
($\Re\{\Omega_\pm\}$); and in solid red, the maxima in the spectra
given by numerical solutions to Eq.~(\ref{eq:TueApr28184248GMT2009}).
By comparing the red and the blue lines, it is clear that both the
repulsion of PL lines (cases c and d) and their crossing (cases a and
b) can appear both in SC (b and c) or WC (a and~d).

At resonance, where the spectrum is symmetric with respect to the
central frequency~$\omega_a$, Eq.~(\ref{eq:TueApr28184248GMT2009}) can
be solved in terms of radicals. $\omega_a$ is always a solution,
corresponding to the minimum in the doublet case and to the maximum in
the singlet case. The other two possible real solutions give the
expression for the observed splitting, given by, in both SC or WC:
\begin{equation}
  \label{eq:MonApr6124752BST2009}
  \Delta \omega_O=2g\Re\Big\{\sqrt{\sqrt{\Big(1+\frac{P_b}{P_a}\Big)^2-4\frac{\Gamma_+}{g}\Big(\frac{\Gamma_b}{2g}+\frac{P_b}{P_a}\frac{\Gamma_-}{g}\Big)}-\frac{P_b}{P_a}-\Big(\frac{\Gamma_b}{2g}\Big)^2}\Big\}\,.
\end{equation}
This should be contrasted with the expression for the DM splitting:
\begin{equation}
  \label{eq:MonApr27215728GMT2009}
  \Delta \omega_\mathrm{DM}=2\Re\{R(\Delta=0)\}=2g\Re\Big\{\sqrt{1-(\Gamma_-/g)^2}\Big\}\,.
\end{equation}
Our main statement in this text is that the counterpart of
Eq.~(\ref{eq:MonApr6124752BST2009}) at nonzero detuning does not exist
in this form and should be computed numerically. This is in stark
contrast with the DM case, Eq.~(\ref{eq:MonApr27215728GMT2009}), which
always has an equally simple expression, that in general does not,
however, describe accurately experimental data.

Expression~(\ref{eq:MonApr6124752BST2009}) can be used to predict the
pumping rates at which the splitting will be more visible for a given
configuration. The counterpart expression for the splitting observed
in the direct exciton emission can be obtained by simply exchanging
indexes~$a\leftrightarrow b$. It is also interesting to note that, for
the integrated spontaneous PL emission (instead of the SS emission),
the formula for the observed splitting is found by only
removing~$P_{a,b}$ from the~$\Gamma$'s and substituting~$P_b/P_a$ by
the ratio of initial state
populations~$\mean{\ud{b}b}^0/\mean{\ud{a}a}^0$ (under the assumption
that initially~$\mean{\ud{a}b}^0=0$).

In conclusion, we have contrasted the crossing and anticrossing of the
PL spectrum lines as detuning is varied, with the actual behavior of
the dressed modes, which truly quantify the strength of the
light-matter coupling. We showed that a careful analysis is required
to properly assess an experiment that is not in very strong
coupling. No analytical formula in radicals of the parameters of the
system exists in the general case, supporting the global fitting of
lineshapes as the most convenient description of experimental data.

\emph{Acknowledgements:} EdV and FPL are grateful to A.~Laucht,
S.~Reitzenstein, M.~Glazov, A.~Poddubny, D. Gerace, A.~Dousse and
V.~Kulakovskii for valuable discussions on this topic at the PLMCN9
conference. AGT acknowledges the support of \emph{La Caixa} and EdV of
a Newton Fellowship.


\begin{thebibliography}{00}

% \bibitem{qdexp} J.P. Reithmaier \emph{et al.}, \emph{Nature} 432, 197
%   (2004); T. Yoshie \emph{et al.}, \emph{Nature} 432, 200 (2004);
%   E. Peter \emph{et al.}, \emph{Phys. Rev. Lett.} 95, 067401 (2005)

\bibitem{qdth} F.P. Laussy \emph{et al.}, \emph{Phys. Rev. Lett.},
  101, 083601 (2008); A. Laucht \emph{et al.}, arXiv:0904.4759;
  N.S. Averkiev \emph{et al.}, \emph{JETP}, 135, 959 (2009).

\bibitem{bosons} F.P. Laussy \emph{et al.}, to be published in
  \emph{Phys. Rev. B}, arXiv:0807.3194.

% \bibitem{hermite} C. Hermite, \emph{Annali di math. pura ed appl.}
%   1, 256-259
%   (1858) %in his publication "Sulla risoluzione delle equazioni del
%   quinto grado."

\end{thebibliography}
\end{document}